\documentclass[11pt]{article}
\usepackage{amsfonts}
\usepackage{amssymb}
\usepackage{amscd}
\usepackage{amsmath}
\usepackage{epsfig}
\usepackage{latexsym}
\usepackage{enumerate}

\newlength{\dinwidth}
\newlength{\dinmargin}
\newtheorem{definition}{Definition}
\newtheorem{theorem}{Theorem}
\newtheorem{proposition}{Proposition}

\newtheorem{lemma}{Lemma}

\def\be{\begin{equation}}
\def\ee{\end{equation}}
\def\ben{\begin{displaymath}}
\def\een{\end{displaymath}}
\def\baa{\begin{eqnarray}}
\def\eaa{\end{eqnarray}}

\def\ba{\begin{array}}
\def\ea{\end{array}}

\def \surf {{\cal L}}
\def\i{{\rm i}}

\def\e{\epsilon}

\def\sig{\sigma}

\def\Abel{{\cal U}}

\def\Fcal{{\cal F}}

\makeatletter
\@addtoreset{equation}{section}
\makeatother

\def\CP1{\mathbb {CP}^1}
\def\C{{\mathbb C}}
\def\Z{{\mathbb Z}}

\def\a{\alpha}
\def\g{\gamma}

\def\b{\beta}

\def\l{\lambda}

\def\phi{\varphi}
\def\th{\theta}

\def\o{\omega}

\def\M{\Lambda}

\def\f{\frac}
\def\la{\label}

\def\p{\partial}

\def\per{\mathbb B}

\begin{document}

\title{On higher genus  Weierstrass sigma-function}
\date{}
\author{D. Korotkin$^1$  \; and V. Shramchenko$^2$ }
\maketitle
\hskip6.0cm {\it Dedicated to the 60th birthday of Boris Anatol'evich Dubrovin}
\vskip2.0cm

\footnotetext[1]{e-mail: korotkin@mathstat.concordia.ca;
address: Department of Mathematics and Statistics, Concordia University
1455 de Maisonneuve Blvd. West, Montreal H3G 1M8  Quebec,  Canada}

\footnotetext[2]{e-mail: vasilisa.shramchenko@usherbrooke.ca; address: D\'epartement de Math\'ematiques, 
Universit\'e de Sherbrooke; 2500 boul. de l'Universit\'e, Sherbrooke J1K 2R1 Qu\'ebec, Canada  }

{\bf Abstract.}
The goal of this paper is to propose a new way to generalize the Weierstrass sigma-function to higher genus Riemann
surfaces. 
Our definition of the odd higher genus sigma-function is based on a generalization of the classical 
 representation of the elliptic sigma-function via
Jacobi theta-function. 
Namely, the odd higher genus  sigma-function $\sigma_{\chi}(u)$ (for $u\in \C^g$) is defined as a product of the 
theta-function with odd half-integer characteristic 
$\beta^{\chi}$, associated with a spin line bundle $\chi$,   an exponent of a certain bilinear form, the determinant of 
a period matrix and a power of the product of all  even theta-constants which are non-vanishing on a given Riemann surface.

We also define an even sigma-function corresponding to an arbitrary even spin structure on $\surf$.
Even 
sigma-functions are constructed as a straightforward analog of a classical formula relating even and odd 
sigma-functions. 
In higher genus the even sigma-functions are
well-defined on the moduli space of Riemann surfaces outside of a subspace defined by 
vanishing of the corresponding even theta-constant.

\tableofcontents

\section{Introduction}

Classically, there exist two main approaches to the theory of elliptic functions: the  Jacobi approach
based on the theory of theta-functions, and the Weierstrass approach.
Each of these two pictures has its advantages and disadvantages when tackling a concrete problem, but essentially they
are completely equivalent to each other. Strangely enough, Jacobi's picture admits a very natural
and well-developed generalization to the higher genus case, while the Weierstrass picture remains essentially 
undeveloped.

Generalization of the Weierstrass sigma-function to higher genus started from works of  
 F.Klein who treated the hyperelliptic case in \cite{Klein1} and Baker \cite{Baker}. In \textsection  27  
of subsequent work \cite{Klein2},  F. Klein 
constructed a sigma-function for an arbitrary curve of genus 3, making the first step beyond the hyperelliptic case.
More recently the subject attracted attention of many researchers 
 (Buchstaber, Leykin, Enolskii, Nakayashiki and others \cite{BLE,BLE1,Naka2008,EEG}) who developed the theory 
of higher genus sigma-function for 
  the so-called $(n,s)$-curves. It turns out that the sigma-functions are a convenient tool in
description of algebro-geometric solutions of integrable systems of KP type as well as in description of Jacobi and Kummer
algebraic varieties.


In this paper we introduce a notion of a higher genus sigma-function for an arbitrary Riemann surface of genus $g$.
We construct a straightforward and natural analogue 
of elliptic sigma-functions
following the classical  formalism of the theory of elliptic functions as close as possible.
Our definition of the higher genus sigma-function resembles the genus three sigma-function of Klein from \cite{Klein2},
although we also use some ingredients of recent works  \cite{BLE,Naka2008}. The main role in our approach 
is played by the matrix of logarithmic derivatives of the product of all non-vanishing theta-constants
with respect to entries of the matrix of $b$-periods. The even sigma-functions defined via our scheme are 
modular invariant; as far as the odd sigma-functions are concerned, we can only claim modular invariance of a
certain power of them.

Our generalization of the notion of a sigma-function to an arbitrary Riemann surface is based on the following expression of the sigma-function in terms of the
Jacobi theta-function $\theta_1$, which is the  elliptic theta-function 
with the odd characteristic $[1/2,1/2]$ 
(\cite{WW}, Section 21.43):
 \be
\sig(u)= \f{2\o_1}{\theta_1'}{\rm exp}\left\{\f{\eta_1}{2\o_1} u^2\right\}
\theta_1\left(\f{u}{2\o_1}{\Big |}\f{\o_2}{\o_1}\right)\;,
\la{siggen1int}
\ee
where the parameters $\eta_1$ and $\o_1$  are related via the equation  
\be
\o_1 \eta_1=-\f{1}{12}\f{\theta_1'''}{\theta_1'}\;.
\la{omeetaint}
\ee
Besides the odd sigma-function (\ref{siggen1int}), in classical theory there exist 3 even sigma-functions
\cite{WW}.

To define the higher genus sigma-functions we start from the following data: a 
 Riemann surface $\surf$ of genus $g$ with a chosen 
canonical basis   cycles $\{a_i,b_i\}_{i=1}^g$, 
a marked point $x_0\in \surf$ with some local parameter $\zeta$,  and an odd non-singular spin line bundle $\chi$.

Using the local parameter $\zeta$   we build a distinguished basis of holomorphic differentials $v_j^0$ which is independent of the choice
of a canonical basis of cycles on $\surf$ (the differentials $v_i^0$ are defined by their local behaviour near $x_0$ as follows:
$v_j^0(x)=(\zeta^{k_j-1}+ O(\zeta^{k_j}))d\zeta$, where $(k_1,\dots,k_g)$ is the  Weierstrass gap sequence at $x_0$).
 The matrices of $a$- and $b$-periods of these differentials 
 we denote by $2\o_1$ and $2\o_2$, respectively. We have $\o_2=\o_1\per$, where $\per$ is the matrix of $b$-periods of $\surf$.
Let us denote by ${\cal S}$ the set of   even characteristics $\beta^\mu$ such that
$\theta[\beta^\mu](0|\per)\neq 0$ (or, equivalently, the set of non-vanishing  theta-constants on $\surf$); by $N$ we denote the number of these characteristics. For a generic curve, according to Th.7 of \cite{farkas}, $N=2^{2g-1}+2^{g-1}$; for a hyperelliptic curve 
$N=\f{1}{2}\left(^{2g+2}_{g+1}\right)$.
The key ingredient of our construction is the product of all non-vanishing  theta-constants, we denote this product
 by $\Fcal$:
\begin{equation*}
\Fcal=\prod_{\beta^\mu\in {\cal S}}\theta[\beta^\mu](0|\per)\;.
\end{equation*}
The following symmetric matrix $\M$ which is proportional to the matrix of derivatives of $\log\Fcal$ with respect to matrix entries of $\per$ plays the main role in the sequel:
\begin{equation*}
\M_{ij}= - \f{\pi \i}{N}\f{\partial \log \Fcal}{\partial \per_{ij}}= -\f{1}{4N}\sum_{\beta^\mu\in {\cal S}}\f{1}{\theta[\beta^\mu](0|\per)}\f{\p^2\theta[\beta^\mu](z|\per)}{\p z_i\p z_j}\Big|_{z=0}\;.
\end{equation*}
 This matrix was used in \cite{Klein2} (see \textsection 25) where it played the same role as here in the construction of sigma-functions in genus $3$.

Now we define  two other matrices, $\eta_1$ and $\eta_2$ as follows:
\be
\eta_1 = (\omega_1^t)^{-1} \M\;,\hskip0.8cm
\eta_2={\o_1^{t}}^{-1}\M\per-\frac{\pi \i}{2}{\o_1^{t}}^{-1}\;.
\la{eta12intr}
\ee

The matrices $\o_{1,2}$ and $\eta_{1,2}$ defined in this way satisfy equations
 $\o_2 \o_1^t = \o_1 \o_2^t$, $\eta_2 \eta_1^t = \eta_1 \eta_2^t$ and the higher genus analogue of the Legendre
relation 
$\eta_1 \o_2^t =\eta_2 \o_1^t+\f{\pi \i}{2} I_g$, where $I_g$ is the $g\times g$ unit matrix. Note that the first relation in (\ref{eta12intr}) gives a natural generalization of (\ref{omeetaint}).

Consider the Abel map $\Abel(x)=\int_{x_0}^x v_i$  on $\surf$. Here $v_i$ are holomorphic 1-forms on $\surf$ normalized via
relations $\int_{a_i}v_j=\delta_{ij}$. Our construction does not depend on the choice of the initial point for the Abel map. We choose this point  to coincide with $x_0$, the point of normalisation for the holomorphic differentials $v^0_i$.  

Let us denote by $D$ the positive divisor of degree $g-1$ corresponding to the odd spin line bundle $\chi$.
Since the divisor $2D$ lies in the canonical class, the vector $\Abel(D)+K^{x_0}$ (here $K^{x_0}$ is the vector of Riemann constants corresponding to the point $x_0$) equals $\per \beta_1+\beta_2$, where $\beta_{1,2}\in 1/2\Z^g$. Denote by  $\beta^\chi$, the
odd  theta-characteristic defined by the vectors 
$[\beta_1,\beta_2]$.

As a straightforward  analogy to (\ref{siggen1int}) we define an odd higher genus  sigma-function 
corresponding to an arbitrary
 non-singular odd spin structure $\chi$  as follows:
\be
\sig_{\chi}(u)= \Fcal^{-3/N} {\rm det}(2\o_1)\, {\rm exp}\left(\f{1}{2} u^t (\eta_1\o_1^{-1}) u\right)
\theta[\beta^{\chi}]((2\o_1)^{-1}u|\per)\;,
\la{sigint}
\ee
where  $\theta[\beta^{\chi}](z|\per)$ is the Riemann theta-function with the odd characteristics $\beta^\chi$.

The function  $\sig_\chi$ is  modular invariant in the following sense: 
under a change of canonical basis of cycles $\sig_\chi^{8 N}$
remains invariant, i.e., $\sig_\chi$ itself can get multiplied by a root of unity of degree  $8N$.
Thus modular properties of $\sig_\chi$ are determined by a homomorphism of the modular group to
the cyclic group ${\mathbb Z}_{8N}$.

 The matrices  $\o_{1,2}$ and $\eta_{1,2}$  depend on the choice of a base point $x_0$ and a local parameter $\zeta$. 
However, the
 sigma-functions corresponding to different choices of a marked point turn out to be equivalent: they coincide up to a linear change of the variable $u$.

In addition to the odd sigma-function $\sig(u)$, Weierstrass introduced 
 three even sigma-functions $\sig_r$, $r=1,2,3$, which are proportional 
to even Jacobi theta-function. The even sigma-functions are expressed in terms of $\sig$ as follows:
\be
\sigma_r (u) =\f{e^{-\eta_r u} \sigma(u+\omega_r)}{\sigma(\omega_r)}\;,
\la{sirint0}
\ee
where $\omega_3=-\omega_1-\omega_2$ and $\eta_3=-\eta_1-\eta_2$.
Applying a higher genus analogue of formula (\ref{sirint0}) to any of the odd sigma-functions 
(\ref{sigint}), we define
a higher genus even sigma-function corresponding to an arbitrary even spin line bundle $\mu$:
\be
\sig_{\mu}(u)=  {\rm exp} \left(\f{1}{2} u^t (\eta_1 \o_1^{-1}) u\right)
\f{\theta[\beta^\mu]((2\o_1)^{-1}u|\per)}{\theta[\beta^\mu](0|\per)} \;.
\la{sievenint}
\ee

The functions  $\sig_{\mu}$  themselves (not their $8N$th power, as in the case of $\sig_\chi$) are invariant under any change of the canonical basis of cycles.
The function $\sig_\mu$ is well-defined for all Riemann surfaces for which
$\theta[\beta^\mu](0|\per)\neq0$.


We find our present definition of the higher genus sigma-function  more natural  than the 
  definitions of this object given in \cite{BLE,BLE1,EEG,Naka2008} and other previous works due to the following reasons.
First, in contrast to these works, we do not make  use of  any concrete realization of a Riemann surface in the form of an algebraic equation. Second, in \cite{BLE,BLE1,EEG,Naka2008}  a higher genus sigma-function was defined only for a class of the so-called $(n,s)$-curves. On such a curve there exists a holomorphic $1$-form with the only 
zero of multiplicity $2g-2$ (see \cite{Naka2008}); therefore
these curves form only a tiny subset (a subspace of codimension $g-2$) in the moduli space of Riemann surfaces.
Third,  the genus one relation (\ref{omeetaint}) was previously not carried over to higher genus (except,
perhaps, the hyperelliptic case where some analog of  (\ref{omeetaint}) was derived in \cite{EHKKLS}; however,
the relation from \cite{EHKKLS} differs from ours). 
The formulas relating odd and even sigma-functions in genus one were also not generalized to higher genus.
Fourth, in previous works (except the hyperelliptic case) the moduli-dependent factor which provides the
modular invariance of (a power of) an odd sigma-function was not introduced.
Finally, our definition of the higher genus sigma-function naturally generalizes the approach of F. Klein \cite{Klein2}
to sigma-functions of non-hyperelliptic genus three Riemann surfaces.

The paper is organized as follows.

 In section \ref{theta} we collect necessary  facts about theta-functions.
In section \ref{Wsigma} we review definitions and properties of elliptic sigma-functions.
In section \ref{auxobj} we introduce a few auxiliary objects and study their transformation properties under the
change of canonical basis of cycles.
In section \ref{sigmafun}  we define the odd sigma-functions for a generic 
Riemann surface  of arbitrary genus and analyse their periodicity and modular properties. 
In section  \ref{sigmaeven} we introduce even sigma-functions in arbitrary genus and study their properties.
In section \ref{invar} we show that the sigma-functions corresponding to a different choice of the base point
and the local parameter are equivalent, i.e., can be obtained from each other by a linear change of variables. 
In Section \ref{sect_onRS} we replace the argument $u$ of the sigma-function by the Abel map of a point on the Riemann surface and thus consider the sigma-function as a function on the surface. 
In Section \ref{sechyper} we use Thomae's formulas to treat  the hyperelliptic case in more detail.

\section{Theta-function: summary}
\la{theta}

Let us recall the definition and properties of the Riemann theta-function.

The genus $g$ theta-function with characteristics  $\left[^\a_\b\right]$ (where $\alpha,\b\in\C^g$), with the matrix of periods $\per$ and an argument $z\in \C^g$  is defined as follows:
\begin{equation*}
\th\left[^\a_\b\right](z|\per)=\sum_{m\in \Z^g} {\rm exp} \{\pi \i ((m+\a)^t)\per (m+\a) + 2\pi \i (m+\a)^t\, (z+\b)\}\;.
\end{equation*}
The theta-function possesses the following quasi-periodicity properties with respect to shifts of its argument by 
the period vectors:
\be
\th\left[^\a_\b\right](z+ {\bf k}_1|\per)=   e^{2\pi \i (\a^t) {\bf k}_1}\th\left[^\a_\b\right](z|\per)  
\la{pertheta1}
\ee
and
\be
\th\left[^\a_\b\right](z+\per {\bf k}_2|\per)= e^{-2\pi \i (\b^t) {\bf k}_2}
e^{-\pi \i (k_2^t)\per k_2 -2\pi \i  (k_2^t)z }  \th\left[^\a_\b\right](z|\per) , 
\la{pertheta2}
\ee
where ${\bf k}_{1,2}\in \Z^g$ are arbitrary.

To describe modular properties of the theta-function consider a symplectic transformation of the basis of cycles on the surface $\surf$
\be
\left(\ba{c} {{\bf b}}^\gamma \\
              {{\bf a}}^\gamma     \ea\right)= \gamma 
\left(\ba{c} {\bf b} \\
              {\bf a}\ea\right)
\la{gamma1}
\ee
defined by the matrix
\be
\gamma=\left(\ba{cc} A   &  B \\
                     C   &   D \ea\right)\;\;\in\;\; Sp(2g,\Z)\;. 
\la{gamma}
\ee
Here ${\bf a} = (a_1,\dots,a_g)^t$ and ${\bf b} = (b_1,\dots,b_g)^t$ are vectors composed of basis cycles.

The corresponding modular transformation of the theta-function looks as follows (see for example 
\cite{Fay73}, page 7):
\be
\th\left[^\a_\b\right]^\g(((C\per +D)^t)^{-1} z| \per^\g)= \xi(\g,\a,\b)\;\{{\rm det} (C\per+D)\}^{1/2} 
e^{\pi \i\left\{ (z^t)(C\per +D)^{-1} Cz\right\}} \; \th\left[^\a_\b\right] (z|\per),
\la{thgam}
\ee
where $\xi(\g,\a,\b)=\rho(\g)\kappa(\g,\a,\b)$; $\rho(\g)$ is a root of unity of degree 8;
\be
\kappa(\gamma, \a, \b) = {\rm exp}
\{ \pi \i [({\a}^t D^t  -  {\b}^t C^t)(- B{\a}+A{\b}+(A(B)^t)_0) - ({\a}^t){\b}]\}\;,
\ee
and
\be
\left[\ba{c} \a \\ \b \ea\right]^\g =\left(\ba{cc} D & -C \\  -B & A \ea \right) 
\left[\ba{c} \a \\ \b \ea\right] + \f{1}{2} \left[\ba{cc}  (C D^t)_0  \\ (A B^t)_0 \ea\right]\;.
\la{transcar1}
\ee
For an arbitrary matrix $M$, the notation $M_0$ is used for the column vector of diagonal entries of $M$.

The transformation of the Riemann matrix $\per$ of $b$-periods is as follows \cite{Fay92}: 
\be
\per^\g=(A\per+B)(C\per+D)^{-1}.
\la{transper}
\ee

\section{Weierstrass sigma-function}
\la{Wsigma}

Let us first briefly discuss the classical  Weierstrass sigma-function from a convenient perspective (see, for example, chapter 20 of \cite{WW}).

Let $\surf$ be a Riemann surface of genus $1$ with an arbitrary (not necessarily normalized)
 holomorphic differential $v^0$ on $\surf$ and a canonical basis  $\{a,b\}$ of $H_1(\surf,\Z)$.
Introduce $a$- and $b$-periods of $v^0$: 
$2\o_1:=\oint_a v^0$ and $2\o_2:=\oint_b v^0$. Choosing $x_0\in \surf$ as the initial point, we can map the surface $\surf$
to the fundamental parallelogram $J(\surf)$ 
with periods $2\o_1$ and $2\o_2$ and identified opposite sides. This map is given by $u(x)=\int_{x_0}^x v^0$; then $u$ can be used as 
a local coordinate on $\surf$.
Now introduce  the Weierstrass
$\wp$-function, which is a double periodic function of $u$ with the periods $2\o_1$ and $2\o_2$ and a second order pole at $0$ such that 
$\wp(u)\sim u^{-2}+o(1)$ as $u\to 0$.
Then the $\zeta$-function is defined via $-\zeta'(u)=\wp(u)$ and $\zeta(u)=u^{-1}+o(1)$ as $u\to 0$, and, finally, the sigma-function is defined by the
equation $\zeta(u)=\sigma'(u)/\sigma(u)$ and $\sigma(u)= u+ o(u)$ as $u\to 0$.

 The sigma-function defined in this way has the following properties which characterize it uniquely:

\begin{enumerate}[A.]
\item
$\sig(u)$ is  holomorphic in the fundamental 
parallelogram with the sides $2\o_1$ and $2\o_2$ and has a simple zero at $u=0$.

\item
$\sig(u)$ satisfies the following periodicity relations:
\begin{equation*}
\sig(u+2\o_1) = -e^{2 \eta_1 (\o_1 +   u) }  \sig(u)\;,\hskip0.7cm
\sig(u+2\o_2) = -e^{2 \eta_2 (\o_2 +   u) }  \sig(u),
\end{equation*} 
where $\eta_1:=\zeta(\o_1)$ and  $\eta_2:=\zeta(\o_2)$; these constants  are related to periods $\o_1$ and $\o_2$ via the Legendre
relation $\eta_1\o_2-\eta_2\o_1=\pi \i/2$ and
\begin{equation*}
\o_1\eta_1=-\f{1}{12}\f{\theta_1'''}{ \theta_1'}\;. 
\end{equation*}

\item
Consider an arbitrary matrix $\g\in SL(2,\Z)$ of the form (\ref{gamma})
 acting on the  periods $\o_{1,2}$  as follows:
 $\o_1^\g= C \o_2+ D\o_1$, $\o_2^\g= A\o_2+ B\o_1$.
Then the sigma-functions corresponding to two sets of periods, coincide:
\begin{equation*}
\sig(u; \o_1,\o_2)=\sig(u;\o^\g_1,\o^\g_2)\;. 
\end{equation*}

\item 
The sigma-function is locally holomorphic as a function of  periods  $\o_1$ and $\o_2$, which play the role of moduli parameters.
Moreover, $\sig(u;\o_1,\o_2)$ neither vanishes nor diverges identically in $u$ for any values of $\o_1$ and $\o_2$ as
long as $\Im(\o_1/\o_1)$ remains positive.
 
\item
Normalization: $\sigma'(0)=1$.

\end{enumerate}

These properties determine the way to generalize the classical sigma-function to an arbitrary genus. However, in higher genus it turns 
out to be impossible to satisfy all of the properties A-E simultaneously. Therefore, we shall keep only the (appropriately reformulated)
 properties A-C as the main principles.  
The property D, as it stands, can not be fulfilled in arbitrary genus. Namely, the higher genus sigma-function
is well-defined and holomorphic with respect to moduli on each stratum of the moduli space where the given set
of theta-constants remains non-vanishing. On the subspace of the moduli space where some of these theta-constants vanish the sigma-function 
becomes singular as a function of moduli, and has to be redefined using the new set of non-vanishing theta-constants.
The property E does not have an obvious natural analog in the  case of a function of several variables; therefore, we are not going to carry it over to the higher genus case.

The odd Weierstrass sigma-function is expressed in terms of the
Jacobi theta-function $\theta_1$, and periods $\o_1$, $\o_2$ and $\eta_1$ (\cite{WW}, Section 21.43) by (\ref{siggen1int}).
 %

This formula 
 is the starting point of our construction of higher genus odd sigma-functions.


To construct even $\sigma$-functions in any genus we shall generalize formula (\ref{sirint0}).
%

\section{Some auxiliary objects}
\la{auxobj}

\subsection{Definitions}

Fix a Riemann surface $\surf$ with a marked point $x_0$ and a chosen local parameter 
$\zeta$ in neighbourhood of $x_0$.

Let
$1=k_1< k_2,\dots, < k_g\leq 2g-1$ be the Weierstrass gap sequence at $x_0$ (if $x_0$ is a 
 non-Weierstrass point, the gap sequence is  $(1,2,\dots,g-1,g)$).

\begin{definition}\la{defv}
The basis of holomorphic differentials $v^0_1,\dots,v^0_g$  is called ``distinguished'' if in a neighbourhood of 
$x_0$ the holomorphic differentials $v^0_i$ have the following expansion in the distinguished local parameter $\zeta$:
\be
v^0_j(x)=(\zeta^{k_j-1}+ O(\zeta^{k_g})) d\zeta\;.
\la{defvjw}
\ee

\end{definition}

The existence of holomorphic differentials with zeros of order (exactly) $k_j-1$ at the Weierstrass point $x_0$ is an
immediate corollary of the Riemann-Roch theorem.
The required structure of higher order terms in the Taylor series can always be achieved by taking linear 
combinations of such differentials.

 Now let us choose some symplectic basis $\{a_i,b_i\}_{i=1}^g$ in $H_1(\surf,\Z)$, and introduce matrices of $a$- and $b$-periods of  $v_i^0$:
\begin{equation*}
2(\o_1)_{ij}=\int_{a_j}  v_i^0   \;,\;\;\;\;
2(\o_2)_{ij}=\int_{b_j}  v_i^0\;. 
\end{equation*}
Then the matrix of $b$-periods of the surface $\surf$ is given by
\be
\per = \o_1^{-1}\o_2 \;.
\la{Bper}
\ee

Denote the set of all non-singular even theta-characteristics $\beta^\mu$ on $\surf$  by ${\cal S}$, 
and their number by $N$; for a generic surface $N=2^{2g-1}+2^{g-1}$. Consider the product of all non-vanishing theta-constants:
\be
\Fcal=\prod_{\beta^\mu\in {\cal S}}\theta[\beta^\mu](0|\per)\;.
\la{Fcaldef}
\ee 

Let us introduce the following $g\times g$ symmetric matrix $\M$:
\be
\M_{ij}= - \f{\pi \i}{N}\f{\partial \log \Fcal}{\partial \per_{ij}} \;,
\la{defM1}
\ee
which, according to the heat equation for theta-function, can be written as
\be
\M_{ij}=-\f{1}{4N}\sum_{\beta^\mu\in {\cal S}}  \f{1}{\theta[\beta^\mu](0|\per) }\f{\p^2\theta[\beta^\mu](z|\per)}{\p z_i\p z_j}\Big|_{z=0}\;.
\la{defM2}
\ee

Let us now define matrices $\eta_1$ and $\eta_2$ as follows:

\be
\eta_1 = (\omega_1^t)^{-1} \M\;,\hskip0.8cm
\eta_2={\o_1^{t}}^{-1}(\M\per-\frac{\pi \i}{2}I_g)\;.
\la{eta12}
\ee

Definition (\ref{eta12}) together with the  symmetry of matrices $\per=\o_1^{-1}\o_2$ and $\M$
 immediately 
imply the following relations:
\begin{equation*}
-\o_2 \o_1^t + \o_1 \o_2^t =0\;,
\end{equation*}
\begin{equation*}
-\eta_2 \eta_1^t + \eta_1 \eta_2^t =0\;,
\end{equation*}
\be
-\eta_2 \o_1^t + \eta_1 \o_2^t =\f{\pi \i}{2} I_g\;.
\la{Leg3}
\ee

Relation (\ref{Leg3}) is a straightforward higher genus analogue of the Legendre relation.

Definition (\ref{eta12}) implies also the following  relation between $\o_{1,2}$ and $\eta_{1,2}$:
\begin{equation}
\label{useless}
 \eta_1^t\omega_2 -\omega_1^t\eta_2  = \frac{\pi \i}{2}I_g. 
\end{equation}

In the sequel we shall make use of the matrix
\be
\eta_1\o_1^{-1} = (\o_1^t)^{-1} \M \o_1^{-1},
\la{eta1ome1}
\ee
which is obviously symmetric as a corollary of the symmetry of $\M$.

\subsection{Transformation properties}

Let us now see how all the matrices defined in Section \ref{auxobj} transform under a symplectic transformation (\ref{gamma1}), (\ref{gamma}) of the canonical basis of cycles on $\surf$.
%
%
\begin{lemma}\la{chaangeM}
Under a change of the canonical homology basis (\ref{gamma1}) the matrix $\M$ transforms as follows:
\begin{equation}
\label{Lambda_transform}
\M^\gamma =  (C\per+D)\M (C\per+D)^t -\frac{\pi \i}{2} C(C\per+D)^t. 
\end{equation}
\end{lemma}
{\it Proof.}
Let us use the formula (\ref{defM1}) for $\M$. Due to (\ref{thgam}) we have
\be
\Fcal^\g=\epsilon \{{\rm det}(C\per +D)\}^{N/2}\Fcal ,
\la{transF}
\ee
where $\epsilon$ is a root of unity of degree $8$,
i.e.,
\be
\log\Fcal^\g=\log\Fcal+ \f{N}{2}\log \{{\rm det}(C\per +D)\} +\log\e \;.
\la{transFcal}
\ee
Using definition (\ref{defM1}) of the matrix $\M$, we get
\be
(\M^\gamma)_{ij} =-\f{\pi \i}{N}\f{\p \Fcal^\g}{\p\per^\g_{ij}}\;.
\la{Mgam00}
\ee
The Riemann matrix $\per$ transforms as in (\ref{transper}) under the change of a symplectic basis. Substituting (\ref{transper}) and (\ref{transFcal}) into (\ref{Mgam00}) and taking into account that
$$\p_{\per_{ij}}\log{\rm det}(C\per+D)=Tr\{[\p_{\per_{ij}}(C\per+D)](C\per+D)^{-1}\}\;,$$
we see that the matrix $\M^\gamma$ is given by (\ref{Lambda_transform}).

Alternatively one can prove the lemma by a straightforward differentiation of (\ref{thgam}) with respect to $z_i$ and $z_j$ and then putting $z=0.$
$\Box$

Now we are in a position to prove transformation formulas for the matrices $\o_{1,2}$ and $\eta_{1,2}$ 
under the change of a canonical basis of cycles:
\begin{lemma}
Under a symplectic transformation (\ref{gamma1}) the matrices  $\o_{1,2}$ and $\eta_{1,2}$  transform as follows:
\be
\o_1^{\g}= \o_2 C^t + \o_1 D^t\;,\;\;\;\;\o_2^{\g}= \o_2 A^t + \o_1 B^t\;,
\la{trans0}
\ee
\begin{equation*}
\eta_1^{\g}= \eta_2 C^t + \eta_1 D^t\;,\;\;\;\;\eta_2^{\g}= \eta_2 A^t + \eta_1 B^t\;.
\end{equation*}
\end{lemma}

{\it Proof.} The transformation of the matrices of periods $\o_1$ and $\o_2$ (\ref{trans0}) follows from the fact
that the choice of holomorphic differentials $v_i^0$ depends only on a marked point $x_0$ and local parameter $\zeta$, and does not depend on the canonical basis of cycles. 

The transformation of $\eta_1$ and $\eta_2$ is immediately implied by their definition
(\ref{eta12}) and the transformation laws for $\o_1$ (given by (\ref{trans0})), $\per$ (given by (\ref{transper})),
$\M$ (given by (\ref{Lambda_transform})) and the following relations for the blocks of a symplectic matrix $\gamma$ (\ref{gamma}): $C^tA=A^tC,$ $D^tB=B^tD,$ $D^tA - B^tC=1$.
$\Box$

\section{Odd sigma-function in higher genus}
\la{sigmafun}

To generalize $\sigma(u)$ to any genus
consider a Riemann surface $\surf$ of genus $g$ and an odd non-degenerate spin structure $\chi$ on $\surf$.
Fix a point $x_0\in \surf$ and a  local parameter $\zeta(x)$ in a neighbourhood of $x_0$. 
Consider the corresponding distinguished basis of holomorphic differentials $v_1^0,\dots,v_g^0$  
and their matrices of periods $2\o_{1}$ and $2\o_{2}$. Introduce matrices  $\eta_{1,2}$ by (\ref{eta12}).

Formula (\ref{siggen1int}) relating the genus one odd sigma-function with the theta-function $\th_1$ has four main ingredients, which will be generalized to any $g>1$ as follows: 
\begin{enumerate}\rm 
\item
The theta-function with an odd half integer characteristic (which is unique in genus 1) $\theta_1({u}/{2\o_1})$
(recall that $\theta_1=-\theta[1/2,1/2]$).

For genus $g$ Riemann surfaces we choose some odd non-singular spin structure $\chi$ and 
(ignoring the minus sign relating $\theta_1$ and $\theta[1/2,1/2]$ since $\theta_1$ enters the 
definition of $\sigma$ twice) 
 replace  $\theta_1({u}/{2\o_1})$  by 
$\theta[\beta^\chi]\left((2\o_1)^{-1} u\right)$, with $u\in \C^g$, where 
$[\beta^\chi]=\left[^{\beta_1}_{\beta_2}\right]$ and the vectors $\beta_{1,2}\in (1/2)\Z^g$ are defined by
\be
\per\beta_1+\beta_2=K^{x_0}+\Abel_{x_0}(D)\;.
\la{defbetachi}
\ee
Here $D$ is the divisor corresponding to the  spin structure $\chi$, $\;\Abel_{x_0}$ is the Abel map and $K^{x_0}$ is the vector of Riemann constants with the base point $x_0.$
\item
The exponent of the expression $({\eta_1}/{2\o_1}) u^2$.
For any $g>1$ this term is replaced by the exponent of the bilinear form $\f{1}{2} u^t (\eta_1 \o_1^{-1}) u$.

\item
The factor $2{\o_1}$, which does not depend on the argument $u$.
For any $g>1$ this factor is replaced by ${\rm det} (2\o_1)$.
 
\item
The factor $\theta_1'$, which equals $\pi \th_2\th_3\th_4$ according to Jacobi's formula. This factor does not depend on the argument $u$, 
but depends on $\per=\o_1^{-1}\o_2$.  
For any $g>1$ we shall replace this factor by $\Fcal^{3/N}$, where $\Fcal$  given by (\ref{Fcaldef})  is the  product  of all 
non-vanishing theta-constants on $\surf$.

\end{enumerate}

\begin{definition}
The  sigma-function corresponding to an odd non-singular spin structure $\chi$ is defined  by the formula
\be
\sig_{\chi}(u)= \Fcal^{-3/N}\, {\rm det}(2\o_1)\, {\rm exp} \left(\f{1}{2} u^t (\eta_1 \o_1^{-1}) u\right)\,
\theta[\beta^\chi]((2\o_1)^{-1}u|\per).
\la{conjsig}
\ee
\end{definition}
Obviously, $\sig_{\chi}(u)$ is an odd function since $\theta[\beta^\chi]((2\o_1)^{-1}u|\per)$ is odd.
Now we are going to study other properties of the odd sigma-function.

\subsection{Periodicity of odd sigma-functions}

Periodicity properties of $ \sig_{\chi}(u)$ (\ref{conjsig}) are given by the following proposition:

\begin{proposition}

The function $\sig_{\chi}(u)$ has the following transformation properties with
respect to shifts of the argument $u$ by lattice vectors $2\o_1{\bf k}_1$ and $2\o_2{\bf k}_2$ for any
${\bf k}_{1,2}\in \Z^g$:
\be
\sig_\chi(u+2\o_1 {\bf k}_1) = e^{2\pi \i (\beta_1^t)k_1}     e^{2( \o_1{\bf k}_1 + u)^t \eta_1{\bf k}_1}  \sig_\chi(u),
\la{perisig1}
\ee
\be
\sig_\chi(u+2\o_2 {\bf k}_2) =e^{2\pi \i (\beta_2^t)k_2} e^{2( \o_2{\bf k}_2 + u)^t \eta_2{\bf k}_2}\sig_\chi(u)
\la{perisig2}
\ee
where $\beta_1$ and $\beta_2$ are vectors forming the non-singular odd characteristic $\beta^\chi$.
\end{proposition}

{\it Proof.} To prove (\ref{perisig1}) we first use the corresponding periodicity property of the theta-function 
(\ref{pertheta1}), which produces the first exponential factor in  (\ref{perisig1}).
 The  multiplier coming from the exponential term in (\ref{conjsig}) 
 equals
$$
{\rm exp}\frac{1}{2}\{(u^t+ 2 {\bf k}_1^t \o_1^t)\eta_1\o_1^{-1}(u+2\o_1  {\bf k}_1) - u^t \eta_1\o_1^{-1} u\},
$$
which can be brought into the form of the second exponential factor in (\ref{perisig1}) by a simple computation using the symmetry of the matrix $\eta_1^t\omega_1=\M$.

The proof of (\ref{perisig2}) is slightly more complicated; besides  (\ref{pertheta2}) it requires also the use of
  relation (\ref{Leg3}). 

Namely, consider the expression
\be
{\rm log}\f{\sig_\chi(u+2\o_2 {\bf k}_2)}{\sig_\chi(u)} .
\la{comtr}\ee
The contribution from the exponential term in (\ref{conjsig}) to this expression is
$$
\frac{1}{2}\{(u^t+ 2 {\bf k}_2^t \o_2^t)\eta_1\o_1^{-1}(u+2\o_2  {\bf k}_2) - u^t \eta_1\o_1^{-1} u\},
$$
which equals
\be
 2{\bf k}_2^t \o_2^t \eta_1 \o_1^{-1} \o_2 {\bf k}_2+2 u^t \eta_1\o_1^{-1}\o_2 {\bf k}_2 \;,
\la{comper1}
\ee
where the symmetry of the matrix $\eta_1\omega_1^{-1}$ (\ref{eta1ome1}) was used.
The contribution from the second exponential term in the transformation of the theta-function (\ref{pertheta2}) to
(\ref{comtr}) equals
\be
-\pi \i {\bf k}_2^t \o_1^{-1}\o_2 {\bf k}_2 - \pi \i {\bf k}_2^t \o_1^{-1} u\;.
\la{contrtheta}
\ee

Now the sum of first terms in (\ref{comper1}) and (\ref{contrtheta}) gives

\begin{equation}
\label{temp1_periodicity}
2 {\bf k}_2^t\left\{\o_2^t \eta_1  - \frac{\pi\i}{2}I_g\right\} \o_1^{-1}\o_2{\bf k}_2 = 2  {\bf k}_2^t \eta_2^t \o_2  {\bf k}_2 = 2  {\bf k}_2^t \o_2^t\eta_2   {\bf k}_2\;,
\end{equation}
where  in the first equality we used the relation (\ref{useless}). In the second equality we used  the symmetry of 
$\o_2^t\eta_2=\per\M\per -\f{\pi \i}{2}\per$.

Now consider the sum of second terms in  (\ref{comper1}) and (\ref{contrtheta}). This sum is equal to
\begin{equation}
\label{temp2_periodicity}
2u^t\left( \eta_1\omega_2^t  - \frac{\pi\i}{2}\right)(\omega_1^t)^{-1} {\bf k}_2 = 2u^t\eta_2 {\bf k}_2,
\end{equation}
where we used the symmetry of matrix  $\per=\o_1^{-1}\o_2$ and (\ref{Leg3}). The sum of expressions (\ref{temp1_periodicity}) and (\ref{temp2_periodicity}) gives the second exponent in (\ref{perisig2}).
$\Box$

\subsection{Transformation of $\sig_\chi$ under change of canonical basis of cycles}

To deduce modular properties of the sigma-function, we need the following lemma.
\begin{lemma}
For  an arbitrary $\g\in Sp(2g,\Z)$ of the form (\ref{gamma}) acting on symplectic homology basis on $\surf$ according to (\ref{gamma1}) the characteristics $(\beta^\chi)^\g$ and $\beta^\chi$ are related by (\ref{transcar1}).
\end{lemma}
{\it Proof.} According to  Lemma 1.5 on p.11 of \cite{Fay92}, the vectors of Riemann constant $K^\g_{x_0}$ and $K_{x_0}$
are related by (modulo the lattice of periods):
\begin{equation*}
K^\g_{x_0}\equiv [(C\per+D)^t]^{-1} K_{x_0}+ \f{1}{2}\per (C D^t)_0  + \f{1}{2} (A B^t)_0,
\end{equation*}
which immediately shows that  $(\beta^\chi)^\g$ and $\beta^\chi$ are related by (\ref{transcar1}).
$\Box$

The modular properties of the odd sigma-function are described in the next theorem.

\begin{theorem}
The function
$\sigma_\chi(u)$ (\ref{conjsig})
is invariant with respect to symplectic transformations up to a possible multiplication with a root of unity of  degree
$8N$.

\end{theorem}
{\it Proof.}
Take an arbitrary $\g\in Sp(2g,\Z)$ of the form (\ref{gamma}) acting on a symplectic homology basis on $\surf$ according to (\ref{gamma1}) and
 consider the ratio of sigma-functions:
\be
\f{\sig_\chi^\g(u)}{\sig_\chi(u)}= e^{\f{1}{2} u^t (\eta^{\g}_1 (\o_1^\g)^{-1}- \eta_1 \o_1^{-1}) u}\left(\f{{\rm det} \o_1^\g}{{\rm det} \o_1}\right)
\left(\f{\Fcal^{\gamma}}{\Fcal}\right)^{-3/N}
\f{\theta[{\beta^\chi}^\g]((2\o^\g_1)^{-1}u|\per^\g)}{\theta[\beta^\chi]((2\o_1)^{-1}u|\per)}.
\la{sisig}
\ee

According to (\ref{thgam}), we have
\be
\f{\theta[{\beta^\chi}^\g]((2\o^\g_1)^{-1}u|\per^\g)}{\theta[\beta^\chi]((2\o_1)^{-1}u|\per)}=
\xi(\g,\a,\b)\{{\rm det} (C\per+D)\}^{1/2}e^{\pi \i\left\{ (z^t)(C\per +D)^{-1} Cz\right\}},
\la{thgam1}
\ee
where  $z=(2\o_1)^{-1}u$.

Look first at the exponential factor in the ratio (\ref{sisig}) composed of the first multiplier in 
(\ref{sisig}), as well as the last term in the transformation of theta-function (\ref{thgam1}). 
It is convenient to use the variable $z=(2\o_1)^{-1}u$. The term in the exponent arising from (\ref{thgam1}) equals
\be
\pi \i (z^t) (C \per +D)^{-1}C z = \pi \i (z^t) (C^t) (\per C^t + D^t) ^{-1} z.
\la{expth}
\ee
On the other hand, the first term in (\ref{sisig}) can be transformed as follows:
\begin{multline*}
2 (z^t)(\o_1)^t\left\{ (\eta_2 C^t +\eta_1 D^t)(\o_2C^t +\o_1 D^t)^{-1} -\eta_1 \o_1^{-1}\right\}\o_1 z
\\
= 2 (z^t)(\o_1)^t\left\{ (\eta_2 C^t +\eta_1 D^t) (\o_1^{-1}\o_2 C^t +D^t)^{-1} -\eta_1\right\} z.
\end{multline*}
Recall now that $\per=\o_1^{-1}\o_2=\o_2^t(\o_1^t)^{-1}$, 
and the above expression rewrites as
\be
2 (z^t)(\o_1)^t\left\{\eta_2 C^t +\eta_1 D^t -\eta_1(\o_2^t(\o_1^t)^{-1} C^t +D^t)\right\} (\per C^t + D^t) ^{-1} z.
\la{prom1}
\ee
Using the relation  $\eta_1 \o_2^t =\eta_2 \o_1^t+\f{\pi \i}{2} I_g$ from (\ref{Leg3}), we rewrite (\ref{prom1}) as
$$
 -\pi \i (z^t)C^t(\per C^t + D^t) ^{-1} z,
$$
which cancels against (\ref{expth}).  Therefore, the exponential term in (\ref{sisig}) is equal to 1.

Consider the term involving ${\rm det} (C\per+D)\}^{1/2}$ in (\ref{sisig}). This term appears from the transformation law (\ref{thgam}) of the
theta-function, as well as from the transformation law (\ref{transF}) of the function $\Fcal$, and the transformation law of 
of ${\rm det} \o_1$; this term cancels out similarly to the genus 1 case.

What remains is a root of unity of eight's degree from the transformation of the theta-function, as well as a root of unity of   degree $8N$ from the transformation of $\Fcal^{3/N}$ which give altogether a root of unity of degree $8N$ in the transformation of $\sig_\chi$.
$\Box$

\section{Even sigma-functions in higher genus}
\la{sigmaeven}

Even elliptic sigma-functions  are given by
\be
\sigma_r (u) =\f{e^{-\eta_r u} \sigma(u+\omega_r)}{\sigma(\omega_r)}
\la{sir}
\ee
for $r=1,2,3$, where $\omega_3=-\omega_1-\omega_2$ and $\eta_3=-\eta_1-\eta_2.$
The enumeration of sigma-functions is  different from theta-functions: $\sigma$ is proportional to $\theta_1$, $\sigma_3$ is proportional to 
$\theta_3$ (i.e. usual theta-function with 0 characteristic), $\sigma_1$ is proportional to $\theta_2$, and $\sigma_2$ to $\theta_4$:
\begin{equation}
\label{sigmas_thetas}
\sigma_1(u) = {\rm exp} \left( \frac{\eta_1u^2}{2\omega_1} \right) \frac{\theta_2(\frac{u}{2\omega_1})}{\theta_2}, \qquad
\sigma_2(u) = {\rm exp} \left( \frac{\eta_1u^2}{2\omega_1} \right) \frac{\theta_4(\frac{u}{2\omega_1})}{\theta_4}, \qquad
\sigma_3(u) = {\rm exp} \left( \frac{\eta_1u^2}{2\omega_1} \right) \frac{\theta_3(\frac{u}{2\omega_1})}{\theta_3},
\end{equation}
where $\theta_k=\theta_k(0)$.

Let us now describe a generalization of the even sigma-functions to the higher genus case.
Denote by $\mu$  an even spin line bundle on the surface $\surf$ (generically such a line bundle does not admit holomorphic sections).
An inspired by (\ref{sigmas_thetas}) definition  of an even sigma-function corresponding to $\mu$ is less ambiguous than in the case of odd sigma-function if we insist on carrying the normalization $\sigma_\mu(0)=1$ to the 
higher genus case.
Denote by $\beta_\mu$ the non-singular even characteristic corresponding (analogously to (\ref{defbetachi})) to the spin structure $\mu$ under some choice of a canonical basis of cycles $\{a_i,b_i\}_{i=1}^g$.

\begin{definition}
The  sigma-function corresponding to an even non-singular spin structure $\mu$ is defined  by the formula
\be
\sig_{\mu}(u)=  {\rm exp} \left(\f{1}{2} u^t (\eta_1 \o_1^{-1}) u\right)
\f{\theta[\beta^\mu]((2\o_1)^{-1}u|\per)}{\theta[\beta^\mu](0|\per)}.
\la{conjsigeven}
\ee
\end{definition}

Periodicity properties of even sigma-functions (\ref{conjsigeven}) coincide with the periodicity properties (\ref{perisig1}), (\ref{perisig2}) of odd sigma-functions,
where the vectors $\b_1$ and $\b_2$ should be substituted by characteristic vectors corresponding to the even spin bundle $\mu$.

Moreover, even sigma-functions $\sig_{\mu}(u)$ are invariant under any change of canonical basis of cycles. This property
is in contrast to the case of odd sigma-functions where only  $\sig_{\chi}^{8N}(u|\per)$ 
can be claimed to be invariant under  any change of canonical basis of cycles.

Finally we have
\begin{equation*}
\sig_{\mu}(0)=1
\end{equation*}
similarly to the genus 1 case.

The even sigma-functions (\ref{conjsigeven}) is well-defined if
\be
\theta[\beta^\mu](0|\per)\neq 0\;.
\la{singeven}
\ee
Therefore, for a generic Riemann surface when none of theta-constants vanishes all even sigma-functions, as well
as all odd sigma-functions, are well-defined. For Riemann surfaces for which relation
(\ref{singeven}) does not hold, the definition of $\sig_\mu$ should be modified, but we do not consider this case here.

\begin{proposition}
Formula (\ref{sir}) relating odd and even sigma functions in the elliptic case admits the following generalization to higher genus:
\begin{equation}
\label{even-odd}
\sigma_\mu(u) = e^{ -u^t(\eta_1 n_1 + \eta_2 n_2)} \frac{\sigma_\chi(u+\omega_1 n_1 + \omega_2 n_2)}{\sigma_\chi(\omega_1 n_1 + \omega_2 n_2)}, 
\end{equation}
where $n_1$ and $n_2$ are integer vectors, $n_1,n_2\in{\mathbb Z}^g$, such that 
\begin{equation*}
\frac{1}{2} \left(\begin{array}{c}n_2 \\n_1\end{array}\right) = \beta^\mu-\beta^\chi. 
\end{equation*}
\end{proposition}
{\it Proof.} The proposition can be proved by the following straightforward calculation. Using definition (\ref{conjsig}) of $\sigma_\chi$, we rewrite (\ref{even-odd}) in the form: 
\begin{multline*}
\sigma_\mu(u) = {\rm exp} \left\{ -u^t(\eta_1n_1 +\eta_2 n_2) + \frac{1}{2} ( u^t + n_1^t \omega_1^t + n_2^t \omega_2^t ) (\eta_1 \omega_1^{-1})  (u+\omega_1n_1 + \omega_2 n_2)  \right. \\
\left. - \frac{1}{2} (n_1^t \omega_1^t + n_2^t \omega_2^t ) (\eta_1 \omega_1^{-1})  (\omega_1n_1 + \omega_2 n_2)    \right\}
\frac{\theta[\beta_\chi] \left( (2\omega_1)^{-1} u + \frac{1}{2} n_1 + \frac{1}{2}\per n_2  \right) }{\theta[\beta_\chi] \left(  \frac{1}{2} n_1 + \frac{1}{2}\per n_2  \right)}.
\end{multline*}
Simplifying the expression in the exponent and taking into account that
\begin{equation*}
\theta[\beta] (z|\per) =  {\rm exp} \{ \pi {\rm i} \beta_1^t\per \beta_1 + 2 \pi {\rm i} (z+\beta_2)^t\beta_1 \}\,
\theta(z+\beta_2 + \per \beta_1|\per) \qquad \mbox{with}\qquad \beta=\left[\begin{array}{c}\beta_1 \\\beta_2\end{array}\right],
\end{equation*}
we get 
\begin{multline*}
\sigma_\mu(u) = {\rm exp} \left\{ -u^t(\eta_1n_1 +\eta_2 n_2) + \frac{1}{2}  u^t(\eta_1 \omega_1^{-1})u
+ u^t(\eta_1 \omega_1^{-1})  (\omega_1n_1 + \omega_2 n_2) - \pi {\rm i} u^t(2\omega^t_1)^{-1}n_2\right\} \\
\times\frac{\theta[\beta^\mu] \left( (2\omega_1)^{-1} u \right)} {\theta[\beta^\mu] ( 0)}. \qquad \qquad \qquad
\end{multline*}
Further simplification of the exponential factor leads to 
\begin{equation*}
\sigma_\mu(u) = {\rm exp} \left\{  \frac{1}{2}  u^t(\eta_1 \omega_1^{-1})u\right\} {\rm exp}\left\{
 u^t\left( -\eta_2\omega_1^t + \eta_1\omega_2^t - \frac{1}{2}\pi{\rm i} I_g\right) (\omega^t_1)^{-1}n_2) \right\} \frac{\theta[\beta^\mu] \left( (2\omega_1)^{-1} u \right)} {\theta[\beta^\mu] ( 0)}, 
\end{equation*}
which due to relation (\ref{Leg3}) for the period matrices $\omega_i$ and $\eta_i$ coincides with (\ref{conjsigeven}). 
$\Box$

\section{Dependence on a choice of a marked point and a local parameter}
\la{invar}

The period matrix $\o_1$ transforms under a change of the marked point $x_0$ and of
the local parameter $\zeta$ which define the distinguished basis of holomorphic differentials. 
This transformation also implies a transformation of 
  the symmetric matrix (\ref{eta1ome1})  $\eta_1\o_1^{-1} = (\o_1^t)^{-1} \M \o_1^{-1}$
of the bilinear form which enters the exponential term in the definition of both odd (\ref{conjsig})
and even (\ref{conjsigeven}) sigma-functions. However, the matrix $\M$ depends only on the Riemann
surface $\surf$ and a choice of a canonical basis of cycles on $\surf$. 

This allows to establish the coincidence of the sigma-functions corresponding to the same 
spin structure on $\surf$ and to different bases of distinguished differentials.

Namely, consider two points $x_0$ and $\tilde{x}_0$ on the surface with local parameters $\zeta$ and $\tilde{\zeta}$ in their neighbourhoods, respectively. Let $2\o_1$ and $2\tilde{\o}_1$ be the matrices of $a$-periods of two sets of distinguished holomorphic differentials $\{v_i^0\}$ and $\{\tilde{v}_i^0\}$ such that the differentials $\{v_i^0\}$ have expansions (\ref{defvjw}) at the point $x_0$ in $\zeta$ and the differentials $\{\tilde{v}_i^0\}$ have expansions (\ref{defvjw}) at the point $\tilde{x}_0$ in $\tilde{\zeta}$. 

Both sets of differentials give bases in the space of holomorphic differentials on $\surf$ and, therefore, are related by a linear transformation: $ (\tilde{v}_1^0, \dots, \tilde{v}_g^0)^t = Q(v_1^0, \dots, v_g^0)^t$ with some matrix $Q$, i.e., 
\begin{equation*}
\tilde{\o}_1=Q \o_1\;.
\end{equation*}

Then the corresponding odd sigma-functions $\sig_\chi(u)$ and $\tilde{\sig}_\chi(\tilde{u})$ (\ref{conjsig}) look as follows:
$$
\sig_\chi(u)= \Fcal^{-3/N}\, {\rm det}(2\o_1)\, {\rm exp} \left(\f{1}{2} u^t (\o_1^t)^{-1} \M \o_1^{-1}  u\right)\,
\theta[\beta^\chi]((2\o_1)^{-1}u|\per)
$$
and
$$
\tilde{\sig}_\chi(\tilde{u})= \Fcal^{-3/N}\, {\rm det}(2\tilde{\o}_1)\, {\rm exp} \left(\f{1}{2} \tilde{u}^t (\tilde{\o}_1^t)^{-1} \M \tilde{\o}_1^{-1}  \tilde{u}\right)\,
\theta[\beta^\chi]((2\tilde{\o}_1)^{-1}\tilde{u}|\per)\;.
$$
Therefore,
\begin{equation*}
\tilde{\sig}_\chi(\tilde{u})= \{{\rm det} Q\}\,\sig_\chi(Q^{-1} \tilde{u}),
\end{equation*}
or 
\begin{equation*}
\tilde{\sig}_\chi(\tilde{u})= \{{\rm det} Q\}\,\sig_\chi(u),
\end{equation*}
if we put $\tilde{u}=Qu$.

The even sigma-functions are related by even simpler transformation (with $\tilde{u}=Qu$):
\begin{equation*}
\tilde{\sig}_\mu(\tilde{u})= \sig_\mu(u).
\end{equation*}

Thus the sigma-functions corresponding to two different sets of distinguished holomorphic differentials are
 essentially equivalent. We notice however that  the matrix
 $Q$ depends
on the moduli of the Riemann surface, and on the choice of base points $x_0$ and $\tilde{x}_0$, and on the choice of local parameters near $x_0$ and $\tilde{x}_0$.

\section{Sigma-function as a function on a Riemann surface}
\label{sect_onRS}

 One can consider the sigma-function as a function of a point on a Riemann surface, just like the Riemann theta-function.
We recall that the Riemann theta-function, $\theta(z)$, for $z=\Abel(x)-\Abel(D_g)-K$, where $D_g$ is a positive non-special divisor of degree $g$,
has $g$ zeros on $\surf$, and these zeros lie at the points of the divisor $D_g$. (Here we put $\Abel=\Abel_{x_0}$ and $K=K^{x_0}$.) 

What is a  natural definition of the  sigma-function on a Riemann surface? Obviously, we should take the sigma-function 
  $\sig_\chi(u)$ constructed above and take $(2\o_1)^{-1}u+\per\beta_1+\beta_2=\Abel(x)-\Abel(D_g)-K$ for some choice of a divisor $D_g$. A natural set of
$g$ points in our construction consists of the base point $x_0$ and the divisor $D$; therefore, we choose $D_g=x_0+D$.
Then, according to the definition (\ref{defbetachi}) of the characteristic $\beta_\chi$, we have $u=2\o_1(\Abel(x)-\Abel(x_0))=2\o_1\Abel(x)$.

Alternatively, we can introduce the ``modified'' Abel map
\begin{equation*}
U(x)=2\o_1\Abel(x)\;.
\end{equation*}
The components of $U(x)$ are given by integrals of the ``distinguished''
differentials $v^0_j$:
\begin{equation*}
U_j(x)=\int_{x_0}^x v^0_j. 
\end{equation*}

Then the odd sigma-function on the Riemann surface is given  by:
\begin{equation*}
\sig_\chi(x):=\sig_\chi(U(x))\;.
\end{equation*}

Alternatively, using the representation (\ref{conjsig}) of  $\sig_\chi$ in terms of the theta-function, we have:
\begin{equation*}
\sig_\chi(x)= {\cal F}^{-3/N}{\rm det}(2\o_1)\, e^{2 \,\Abel^t(x) \, \Lambda \, \Abel(x)}
\theta[\beta^\chi](\Abel(x)|\per). 
\end{equation*}
 
The function $\sig_\chi(x)$ has zeros at the points $x_0$ and $P_1,\dots,P_{g-1}$ (where $D=P_1+\dots+ P_{g-1}$). 
If we divide it by the spinor $h_\chi(x)$, we get a non-single-valued $-1/2$ form
\begin{equation*}
\hat{\sig}_\chi(x)=\f{\sig_\chi(x)}{h_\chi(x)},
\end{equation*}
which has only one zero at $x=x_0$. This object is similar to the usual prime-form
\begin{equation*}
E(x,x_0)=\f{\theta[\beta^\chi](\Abel(x)-\Abel(x_0))}{h_\chi(x)\,h_\chi(x_0)}\;.
\end{equation*}
The difference is that $ \hat{\sig}_\chi(x)$ is a $-1/2$-form with respect to $x$, and  $x_0$ plays the role of parameter.
In this sense $  \hat{\sig}_\chi(x)$ is similar to the $-1/2$-form
\begin{equation*}
e(x):=E(x,x_0)h_\chi(x_0)=\f{\theta[\beta^\chi](\Abel(x)-\Abel(x_0))}{h_\chi(x)}\;.
\end{equation*}

In contrast to $e(x)$, which transforms non-trivially under modular transformations, our $-1/2$-form, being taken to the power $8N$,
is modular-invariant.

Note also that the second derivative of the logarithm of $\sigma_{\chi}(x-y)$ for two points $x,y\in\surf$
is the following symmetric bidifferential (a differential on $\surf\times\surf$) holomorphic everywhere except for a pole of order $2$ at the diagonal $x=y:$
\begin{equation}
\label{bidiff}
d_x d_y {\rm log}\sigma_\chi(x-y) = W(x,y) -4{\bf v}^t(x)\,\Lambda\, {\bf v}(y)\;. 
\end{equation}
Here ${\bf v} = (v_1,\dots,v_g)^t$ is the vector of holomorphic 1-forms on $\surf$ normalized via the relations $\int_{a_i}v_j=\delta_{ij}$; and $W(x,y):=d_xd_y\log E(x,y)$ is the symmetric bidifferential with a double pole along $x=y$ and holomorphic everywhere else normalized via $\int_{a_i}W(x,y)=0$ for any $i=1,\dots,g$, the integration being taken with respect to any of the arguments. 

The bidifferential 
\be
W_{Klein}(x,y)=W(x,y) -4{\bf v}^t(x)\,\Lambda\, {\bf v}(y)
\la{WKlein}
\ee
from the right-hand side of (\ref{bidiff})
was introduced by F. Klein in \cite{Klein2}  (see also discussion in J. Fay's book \cite{Fay73} p. 22). The bidifferential $W_{Klein}$ is
 independent  of the choice of homology basis defining the bidifferential 
$W(x,y)$, the holomorphic differentials $v_i$ and the Riemann matrix $\per$. This independence can be derived from the transformation (\ref{Lambda_transform}) of the matrix $\M$ and the following transformation of the bidifferential $W$ under a change (\ref{gamma1}) of the canonical homology basis (see \cite{Fay92}, p. 10): 
\begin{equation*}
W^\gamma(x,y) = W(x,y) - 2\pi {\rm i} {\bf v}^t(x) (C\per +D)^{-1} C{\bf v}(y).
\end{equation*}


\section{Hyperelliptic sigma-functions}
\la{sechyper}

Here we consider our general construction presented above for the subspaces of hyperelliptic Riemann surfaces, i.e.,
Riemann surfaces possessing a meromorphic function of degree $2$. 
Any such Riemann surface $\surf$ is biholomorphically equivalent to an algebraic curve of the form
\begin{equation*}
\nu^2=\prod_{j=1}^{2g+2} (\l-\l_j).
\end{equation*}
For definiteness choose the canonical basis of cycles  $\{a_i,b_i\}_{i=1}^g$ on $\surf$ in the same way as in  Chapter III, \textsection 8 of \cite{Tata}.
The point $x_0$ entering our construction can be chosen to coincide with $\infty^{(1)}$ (a point corresponding to $\l=\infty$ where
$\nu\sim\l^{g+1}$) and the local parameter $\zeta = 1/\lambda$. 

The basis of holomorphic differentials $v_i^0$ can in this case be chosen as follows:
\be
v_i^0=-\f{\l^{g-i}d\l}{\nu}, \qquad i=1,\dots, g. 
\ee
The corresponding matrix of periods  is $(2\o_1)_{ij}=\int_{a_i} v_j^0$, and the other matrices 
$\o_2$ and $\eta_{1,2}$ are given by (\ref{Bper}), (\ref{eta12}).

The non-vanishing theta-constants in the hyperelliptic  case correspond to even
characteristics $\beta^\mu=[\beta_1^\mu,\beta_2^\mu]$ which can be constructed via partition 
of the set $S=\{1,\dots,2g+2\}$ into two  subsets: $T=\{{i_1},\dots,{i_{g+1}}\}$ and  $S\setminus T=\{{j_1},\dots,{j_{g+1}}\}$ (see \cite{Tata}); therefore, $N=\f{1}{2}\left(^{2g+2}_{g+1}\right)$. Then the corresponding theta-constant is given by Thomae's formula
\be
\theta[\beta^\mu]^4(0|\per)=\pm {\rm det}^2\,(2\o_1)\prod_{i,j\in T, \;i<j}(\l_i-\l_j)\,\prod_{i,j\not\in T,\; i<j}(\l_i-\l_j).
\la{Thomae}
\ee  
Therefore
\begin{equation*}
\Fcal=\prod_{\beta^\mu }\theta[\beta^\mu](0|\per)=\epsilon [{\rm det}\,(2\o_1)]^{N/2}
\prod_{i,j=1,\,i< j}^{2g+2}(\l_i-\l_j)^{\frac{Ng}{4(2g+1)}},
\end{equation*}
where $\e^8=1$.


 Thus we can transform the definition of the 
odd sigma-function (\ref{conjsig}) to the  form
\begin{equation*}
\sig_\chi(u)=
({\rm det (2\o_1)})^{-1/2}\prod_{i<j} (\l_i-\l_j)^{-3/8}\,{\rm exp} \left(\f{1}{2} u^t (\eta_1 \o_1^{-1}) u\right)
\theta[\beta^\chi]((2\o_1)^{-1}u|\per).
\end{equation*}


 The even hyperelliptic  sigma-function $\sig_\mu$ (\ref{conjsigeven}) can be rewritten 
as follows using (\ref{Thomae}):
\begin{equation*}
\sig_{\mu}(u)= \e\; {\rm exp} \left(\f{1}{2} u^t (\eta_1 \o_1^{-1}) u\right)
\f{\{{\rm det}\,(2\o_1)\}^{-1/2}\theta[\beta^\mu]((2\o_1)^{-1}u|\per)}{\prod_{i<j; i,j\in T}(\l_i-\l_j)^{1/4}\,\prod_{i<j; i,j \not\in T}(\l_i-\l_j)^{1/4}},
\end{equation*}
where $\e$ is a root of unity of degree $8$ which has to be chosen to provide the normalization $\sig_{\mu}(0)=1$.

We see that on the space of hyperelliptic curves the sigma-functions with even 
characteristics are always well-defined as long as the
curve remains non-degenerate.

\section{Open questions}

There are several interesting problems related to results of this
work.  First, the role of sigma-functions in the theory of integrable
systems should be further clarified; results of previous works
\cite{BLE1} suggest that, due to their modular invariance, the sigma-functions might have various
advantages in describing algebro-geometric solutions of integrable
systems in comparison to theta-functions. The issue of modular
invariance should also be resolved completely. Namely,
according to our present results, the sigma-function is invariant under modular
transformations up to multiplication with certain roots of unity.  In particular,
 our present framework allows genus one sigma-function
to  be multiplied with a root of unity of degree 24 under modular transformations, while it is
well-known that this root of unity is in fact absent. 
Therefore, one may wonder whether it should be possible to better control this root of unity 
in the 
higer genus case. This issue might be quite subtle, since even the explicit
computation of the transformation of Dedekind eta-function under the
action of the modular group is rather non-trivial (see \cite{Atiyah}
for the review of this topic).

In the case of Riemann surfaces admitting a holomorphic
differential with zero of the highest order ($2g-2$), and in particular
for the case of the $(n,s)$ curves studied in previous works, it is desirable
to establish an explicit relationship of our construction
with the previous ones. In particular, one could expect that the 
Taylor expansion of the sigma-functions near zero should
involve an appropriate Schur polynomial, similarly to \cite{BLE2,HE}.

{\bf Acknowledgements.} We thank V. Enolskii and A. Kokotov for important discussions. 
We are grateful to S. Grushevsky for attracting our attention to \cite{farkas}.
We are grateful to  the anonymous referee for useful comments.
DK thanks
Max-Planck-Institut f\"ur Mathematik in Bonn where the main part of this work
was completed for support, hospitality and excellent working conditions.
 The work of DK was partially supported by Concordia
Research Chair grant, NSERC, NATEQ and Max-Planck Society.
The work of VS was supported by FQRNT, NSERC and the University of Sherbrooke.

\end{document}